\documentclass[10pt,twocolumn,journal,compsoc]{IEEEtran}
\usepackage{ifpdf}

\ifCLASSOPTIONcompsoc
  \usepackage[nocompress]{cite}
\else
  \usepackage{cite}
\fi
 
\ifCLASSINFOpdf
  \usepackage[pdftex]{graphicx}
  \DeclareGraphicsExtensions{.pdf,.jpeg,.png}
\else
  \usepackage{graphicx}
  \DeclareGraphicsExtensions{.pdf,.jpeg,.png}
\fi

\usepackage[nolist]{acronym}
\usepackage{multirow}
\usepackage{graphicx}
\usepackage{caption}
\usepackage{subcaption}
\usepackage{color}
\usepackage{enumitem}
\usepackage{hyperref}
\usepackage{upgreek}

\def\mPMU/{$\upmu$PMU}

\acrodef{ARP}{Access Reservation Protocol}
\acrodef{UE}{User Entitity}
\acrodef{RAO}{Random Access Opportunity}
\acrodef{MTC}{Machine Type Communication}
\acrodef{TTI}{Transmission Time Interval}
\acrodef{DSO}{Distribution System Operator}
\acrodef{ICT}{Information and Communications Technology}
\acrodef{LTE}{Long Term Evolution}
\acrodef{DER}{Distributed Energy Resources}
\acrodef{XMPP}{Extensible Messaging and Presence Protocol}
\acrodef{MQTT}{Message Queuing Telemetry Transport}
\acrodef{UMA}{User-Managed Access}
\acrodef{TLS}{Transport Layer Security}
\acrodef{PKI}{Public Key Infrastructure}
\acrodef{PMU}{Phasor Measurement Unit}
\acrodef{PMC}{Power Measurements and Control}
\acrodef{DSSE}{Distribution System State Estimation}

\begin{document}
%

\title{Secure Real-Time Monitoring and Management\\ of Smart Distribution Grid\\ using Shared Cellular Networks}

\author{
  \IEEEauthorblockN{%
  Jimmy~J.~Nielsen\IEEEauthorrefmark{1}, %
  Herv\'e~Ganem\IEEEauthorrefmark{2}, %
  Ljupco~Jorguseski\IEEEauthorrefmark{3}, %
  Kemal~Alic\IEEEauthorrefmark{4}, %
  Miha~Smolnikar\IEEEauthorrefmark{4}, %
  Ziming~Zhu\IEEEauthorrefmark{5}, %
  Nuno~K.~Pratas\IEEEauthorrefmark{1}, %
  Michal~Golinski\IEEEauthorrefmark{3}, %
  Haibin~Zhang\IEEEauthorrefmark{3}, %
  Urban~Kuhar\IEEEauthorrefmark{4}, %
  Zhong~Fan\IEEEauthorrefmark{5}, %
  Ales~Svigelj\IEEEauthorrefmark{4}}\\
\vspace{6pt}
\IEEEauthorblockA{\IEEEauthorrefmark{1}Department of Electronic Systems, Aalborg University, Fredrik Bajers vej 7, 9220 Aalborg, Denmark\\
\IEEEauthorrefmark{2}Gemalto, 6 rue de la Verrerie, Meudon 92197, France\\
\IEEEauthorrefmark{3}TNO, Department of Network Technology, Anna van Buerenplein 1, 2595 DA Den Haag, The Netherlands\\
\IEEEauthorrefmark{4}Jozef  Stefan  Institute, Ljubljana SI-1000, Slovenia\\
\IEEEauthorrefmark{5}Toshiba Research Europe Ltd., Telecommunications Research Laboratory, Bristol, BS1 4ND, UK\\
\vspace{6pt}
Emails: %
jjn@es.aau.dk, %
herve.ganem@gemalto.com, %
ljupco.jorguseski@tno.nl, %
kemal.alic@ijs.si, %
miha.smolnikar@ijs.si, %
ziming.zhu@toshiba-trel.com, %
nup@es.aau.dk, %
michal.golinski@tno.nl, %
haibin.zhang@tno.nl, %
urban.kuhar@ijs.si, %
zhong.fan@toshiba-trel.com, %
ales.svigelj@ijs.si}
}
\maketitle

\begin{abstract}
The electricity production and distribution is facing two major changes. First, the production is shifting from classical energy sources such as coal and nuclear power towards renewable resources such as solar and wind. Secondly, the consumption in the low voltage grid is expected to grow significantly due to expected introduction of electrical vehicles. 
The first step towards more efficient operational capabilities is to introduce an observability of the distribution system and allow for leveraging the flexibility of end connection points with manageable consumption, generation and storage capabilities. Thanks to the advanced measurement devices, management framework, and secure communication infrastructure developed in the FP7 SUNSEED project, the \acf{DSO} now has full observability of the energy flows at the medium/low voltage grid. Furthermore, the prosumers are able to participate pro-actively and coordinate with the \ac{DSO} and other stakeholders in the grid. The monitoring and management functionalities have strong requirements to the communication latency, reliability and security. This paper presents novel solutions and analyses of these aspects for the SUNSEED scenario, where the smart grid ICT solutions are provided through shared cellular LTE networks.
\end{abstract}

\IEEEpeerreviewmaketitle

\section{Introduction}
\label{sec:introduction}
In industrialized countries typical consumers are expected to become electricity producers due to the ongoing widespread deployment of distributed generation and energy storage elements, commonly called Distributed Energy Resources (DER). A consumer that produces energy is called a prosumer. In addition to the introduction of DERs, gradual replacement of conventional internal combustion engine vehicles with Electrical Vehicles (EVs) is expected in the near future, causing a significant increase in the load on the power grid. By using only conventional grid management systems the electrical grid capacity is under question.
Reinforcing the grid all the way to the user is an option, though it is expensive, especially when other more convenient and cheaper alternatives are on the horizon. The shift from a mainly unidirectional power flows towards a fully bidirectional paradigm can be used as an advantage, allowing installation of additional DERs within existing infrastructure. However, this requires a precise monitoring of the distribution grid that provides reliable and accurate information on its status to enable dynamic grid management of the future \cite{von2014every}.  

While the benefits and the necessity of a smart distribution grid are clear, the communication solution supporting it is not straightforward. Today, \acfp{DSO} are increasingly using IEC 61850 based communications for high-level monitoring, management and control on high-speed LAN/optical fiber networks \cite{bush2014smart}. However, when the scope is extended downwards to low voltage infrastructure, the availability of such high-end communication solutions is usually not anticipated.
Within the SUNSEED project\footnote{\url{www.sunseed-fp7.eu}}, a promising approach is considered where the already deployed cellular networks (primarily LTE) are used to provide the smart distribution grid communication infrastructure. 
In this paper the focus is on the security framework and network performance requirements to enable the incorporation of various measurement and control devices, which together allow for the establishment of grid management services based on time and privacy sensitive data.

The specific contributions of this paper relate to the smart grid services introduced in the following section. Thereafter, we present the requirements, design choices and proposed solutions for the smart grid communication and security architecture. Next, we consider the performance of shared cellular LTE networks as a part of a smart grid system. Specifically, we study the achievable latency and reliability of the LTE based smart grid communication.
Finally, we summarize our findings and outline the future steps of the SUNSEED project, namely with respect to the large scale field trial that will be deployed until 4th quarter of 2016.

\section{Smart Grid Management Services}\label{sec:smart_grid_monitoring}

\subsection{State Estimation}
To achieve reliable and accurate knowledge on the grid condition, the \ac{DSSE} is of key importance. The benefit of the state estimation is that it can take into account all types of available measurements, thus reducing the investment costs into required measurement infrastructure. 
Further, DSSE provides estimation of grid state also on the grid nodes where measurement devices are not located.
As the measurement locations are placed all the way down to the prosumer level, the shared cellular networks seem to serve as an efficient and viable solution for communicating between measurement devices and the back-end system. In general the DSSE performance depends on location density, type, accuracy, and reporting interval of the available measurement infrastructure in the grid, such as:
\begin{figure*}[tb]
\centering
    \includegraphics[width=0.75\linewidth]{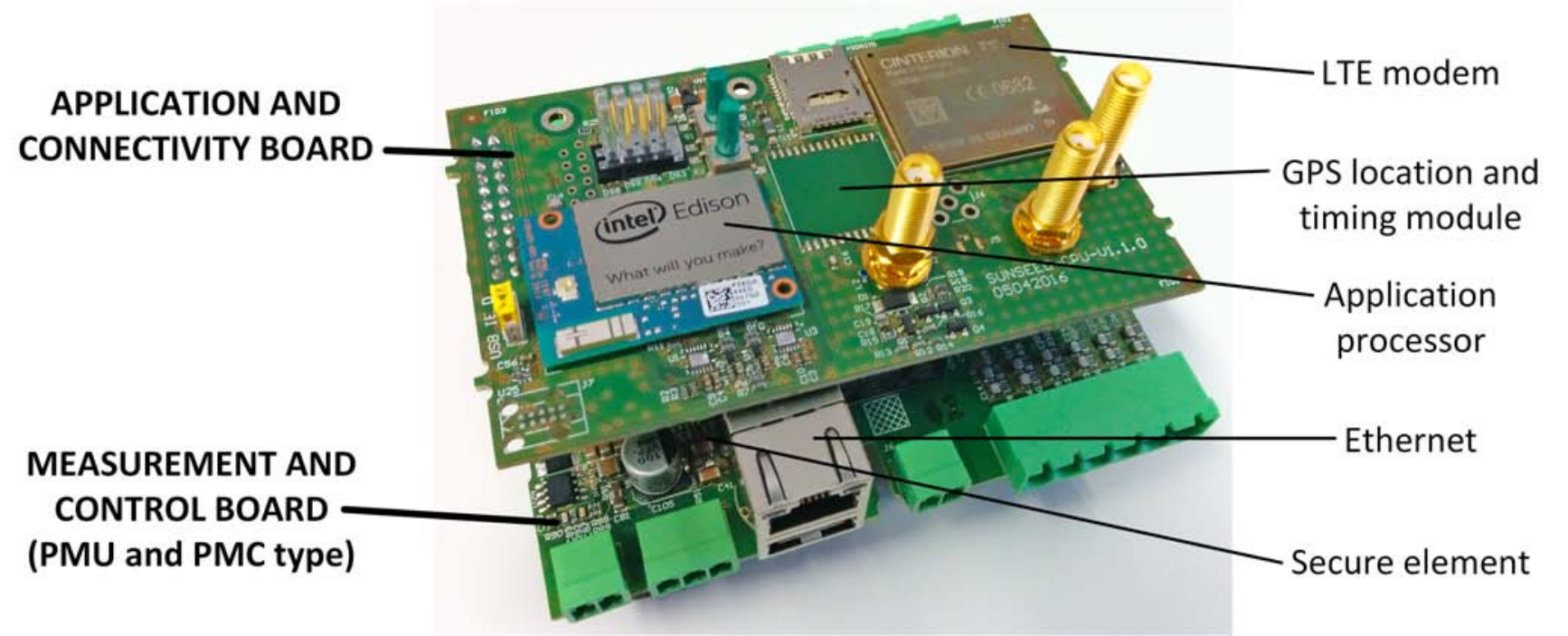}
    \caption{Illustration of \mPMU/ that has been developed for the SUNSEED project.}
    \label{fig:pmu-pmc}
\end{figure*}
\begin{description} 
\item [\acp{PMU}] are dedicated devices with common time reference provided by a very high precision clock, which allow for time-synchronized phasor (that is, synchrophasor) estimations at different locations. Combining high precision and high sampling rate (up to 50/60 Hz) measurements of voltage or current phasors on all 3 phases from multiple PMUs allow for a comprehensive view on the state of the entire grid interconnection.
Fig.~\ref{fig:pmu-pmc} depicts a fully embedded micro PMU (\mPMU/) prototyped within the SUNSEED project that enables 3-phase voltage/current synchrophasor measurements at medium/low voltage of the distribution grid. Besides dedicated measurement circuitry and signal processing it also features a Linux-enabled application processor, LTE, Ethernet and low-power radio communication interfaces, secure element, and a GPS-based reference clock. 
\item [\ac{PMC} devices] allow for 3-phase power quality measurements (such as real/reactive/apparent power, frequency, voltage, current, total harmonic distortion) and control of end connection points (via on/off relays or serial line protocol). Within the SUNSEED project the devices were designed in exactly the same form factor as the \mPMU/, reusing the application and connectivity boards and introducing a measurement and control board. A 1 s reporting period was considered for devices deployed at major grid buses and important prosumer locations to support state estimation, while a request-response mechanism was considered to support Demand-Response services described in section~\ref{sec:dem_resp}. 
\item [Smart Meters (SMs)] for standard billing measurement, assumed to be deployed at each prosumer. Based on 1 min or 15 min reporting interval. In the future, the SMs may be used for power measurements as well, however this requires lower reporting intervals, e.g., down to 1~s like the PMC devices.
\end{description}

Regardless of the challenges that need to be addressed the main benefits of making the grid highly observable can be summarised as follows:
\begin{itemize}
  \item Disturbances on lower voltage level can be locally detected, cleared, and eliminated before they affect other parts of the system.
  \item \acp{DSO} will be able to identify grid model deficits, and allow them to construct accurate models suitable for detailed analysis and planning work.
 \item \ac{DSO} will be able to analyse how installed and planned generation will affect the grid, enabling short, medium and long term planning.
 \item Continuous grid observation will pave the way for the real-time grid control.
\end{itemize}

\subsection{Demand-Response}\label{sec:dem_resp}
The smart distribution grid enables advanced features in demand monitoring, analysis and response. Importantly, the monitoring and control activities will not only reside in operation centers but can also be distributed across the whole grid by enabling control of consumption and production flexibility in consumer and prosumer locations. Having both sides of the distribution grid participating in demand-response can lead to a win-win outcome. The \ac{DSO} benefits from efficient network control and the consumers benefit from optimised use of energy. For example, consumers at the demand side will be able to manage their own consumption by changing the normal electricity consumption patterns over time. Both centralised and decentralised approaches are seen in the literature \cite{ozturk2013intelligent,zhu2015game}. Decentralised techniques usually have reduced computational complexity, however undesired communication overhead may be expected. We notice that the quality of service (QoS) requirement for demand-response is relatively relaxed compared to DSSE, with the measurements interarrival time in the order of minutes to hours, and the data transmission latency requirements around 1~s \cite{osg2013report}. The communication burden can be further reduced with the use of approximated information in the neighbourhood-wide consumption scheduling as proposed in \cite{zhu2016efficient}.
In this way, the scheduling is done by the individual consumers while their actual consumption is observed by the SMs and the PMC units.

\section{Communication and Security Architecture} \label{sec:security_architecture}
The role of the security framework is to protect the smart grid assets against unfriendly attacks. An assessment of potential attacks scenarios has resulted in the identification of four high level security objectives:
\begin{itemize}
  \item Insure availability of the services offered by the smart grid (resilience to cyberattacks to insure functionality).
  \item Insure privacy of communications within the smart grid (avoid spying).
  \item Prevent damage to equipment or infrastructures (resilience to cyberattacks to insure equipment or infrastructures safety).
  \item Avoid fraud (for specific equipment located in subscribers premises).
\end{itemize}
The attacker profiles taken into account are typically:
\begin{itemize}
  \item A cyberterrorist trying to gain information, disrupt  the functioning of the SUNSEED services or lead to a malfunction of the infrastructures by compromising  either  a device,  a communication link or  a cloud platform. 
  \item A subscriber trying to alter the functioning of the smart grid services  to lower its costs  or increase its revenues.
\end{itemize}

The  protection of data communications taking place in the smart grid system is key to meet the security objectives, and different levels of protection may coexist at different levels of the communication protocol stack:
\begin{description}
  \item [Network access level security] targets the protection of the access network. This includes for example the use of SIM cards to authenticate with 3G or 4G wireless networks but also the deployment of VPN, firewalls, etc. 
  \item [Transport level security] primarily aims at protecting point-to-point data communication between two communicating nodes. With communications being IP based, this type of data protection is largely independent from the communication channel (3G, 4G, Wi-Fi, PLC, ...) and is in most cases focused on protection of TCP communication.
  \item [Application level security] addressing the protection of the payload carrying the applications data as described below.
\end{description}

From the security standpoint, the primary high level goal of the architecture described here is to provide end-to-end data protection at the transport level for communication between the monitoring devices and the various smart grid applications and services that access this data. 
End-to-end security in its simplest form consists in implementing point-to-point security, typically at the transport level for every segment in the communication path from source to destination. Every such segment is therefore protected using different credentials and a rekeying operation is typically needed at each transiting node, which should therefore be trusted.
Optionally, it should be possible to add an extra layer of security enabling the ciphering of application data all the way from source to destination (using a group traffic encryption key). In this case the data will remain opaque while moving through transit nodes, possibly reducing the trust level required for these nodes. 

\begin{figure*}[bt]
  \begin{center}
    \includegraphics[width=\linewidth]{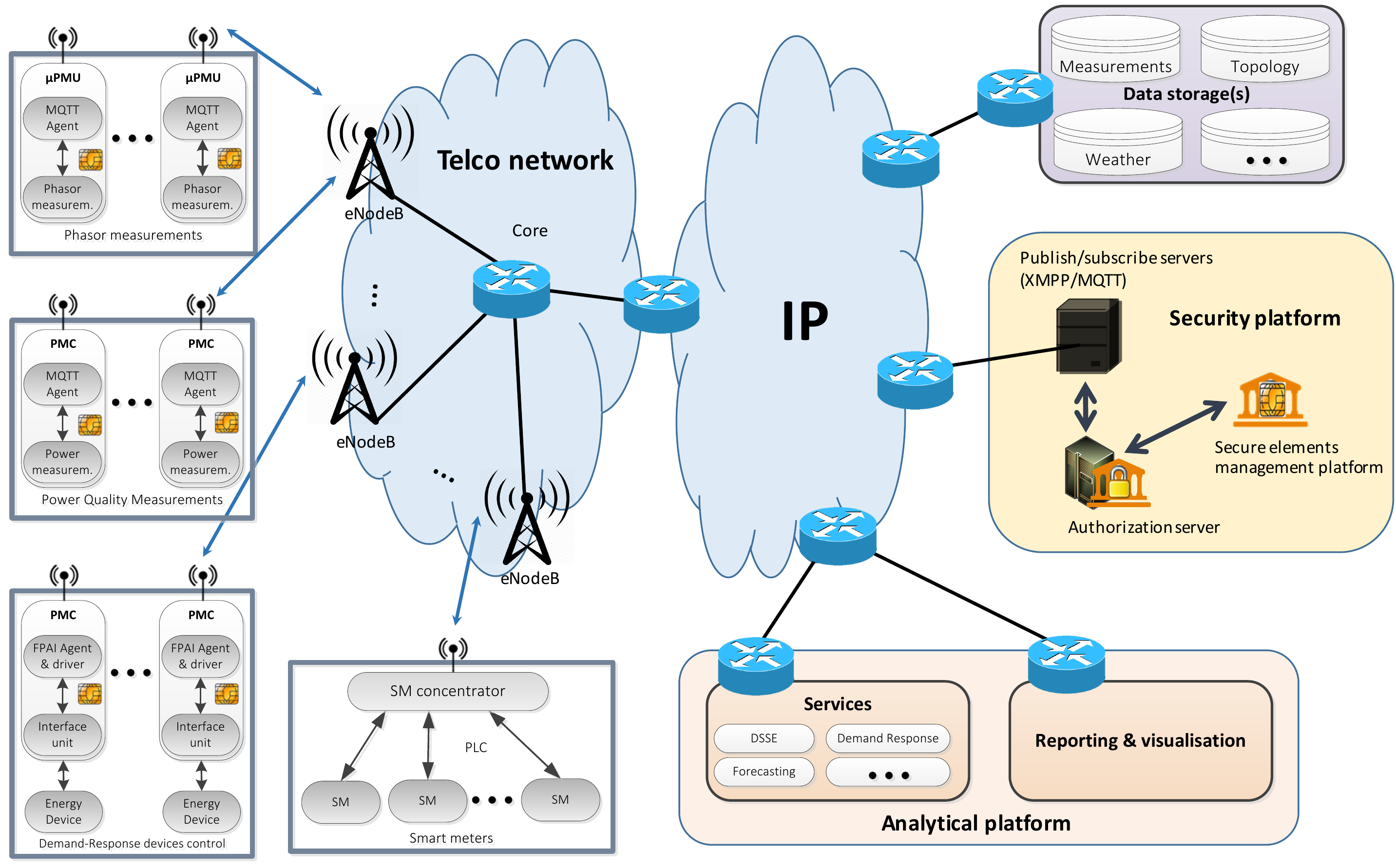}
  \end{center}
  \caption{SUNSEED communication and security architecture.}
  \label{fig:communication_architecture}
\end{figure*}

Fig.~\ref{fig:communication_architecture} describes the proposed communication and security architecture. Data originating from \mPMU/, PMC or SM devices is published to a data sharing platform, possibly transiting through a gateway or aggregator using popular publish/subscribe protocols such as \ac{XMPP} \cite{hardt2012oauth} or \ac{MQTT} \cite{maler2015user}. At the other end of the chain, applications subscribe to the published data to perform specific tasks upon receiving the data. 

Another security requirement addressed by this architecture relates to the capability to manage the authorization rights defining how software entities may interact and exchange data. For optimal consistency and to minimize the risk of security holes resulting from human errors, this is best achieved using a centralized management interface avoiding the need to configure access rights in a disparate way in many different platforms. On that aspect, the choice was made to use a standalone authorization server supporting a delegated authorization scheme. The implementation is based on the \ac{UMA} \cite{hardt2012oauth} profile of the widely used Oauth \cite{hardt2012oauth} protocol. The authorization server is, in this context, the place holding the description of the access control policies to all access controlled ''resources'' in the ecosystem. For example \mPMU/ devices may publish their data on specific information topics using the MQTT protocol, and it is certainly necessary to define which software entity may publish or subscribe on a specific topic. The ''information topic'' is then modeled as an ''access controlled resource''. The delegated authorization scheme, insures a dynamic life cycle management of the access controlled resources in the authorization server, by enabling the resources' servers (for example the publish/subscribe brokers), in charge of enforcing access control, to register and delegate the management of access control for their resources to the authorization server via the UMA REST API. This delegated authorization scheme is one of the two main structural security choices made in the proposed architecture. It opens the possibility to achieve a centralized access control management for resources located in many dispersed heterogeneous platforms. As a result, client applications need only to authenticate with the authorization server to dynamically receive credentials granting them access to multiple heterogeneous platforms. This results not only in enhanced security, but also in a great simplification of authentication and authorization management. 

 The Authorization rights granted to requesting clients are negotiated using the Oauth protocol and materialized as etokens possibly carrying credentials. The idea is to distribute dynamically to every client application (located in the \mPMU/ or PMC devices or in remote cloud servers) the set of credentials required to perform the tasks it should perform with a given work flow scenario. For example, a client application implementing a historical data archive may need to subscribe to the \mPMU/ information topics via the MQTT protocol and then store the received data in an archival database. This client application should then dynamically receive credentials enabling both the reception of information and its storage in the archival database.

Another structural security choice relates to the decision to protect the credentials stored in the \mPMU/ and PMC devices via the use of embedded secure elements, similar to the ones used for manufacturing embedded UICCs \cite{etsi2015requirements} in cellular IoT devices. Commercially available secure elements provide a credible protection for credentials stored inside their memory, which are meant to be used in place and cannot be read back, thereby significantly complicating the setup of attacks involving credential stealing and/or device cloning.

 The protection of communications relies upon the use of the \ac{TLS} protocol involving both \ac{PKI} clients' and servers' certificates. Client certificates are dynamically generated in an initial security bootstrapping process between the \mPMU/ and PMC devices and the authorization server. On the device, the PKI private and public keys are dynamically generated within the secure element. 
 While the public key is sent out and serves as the starting point to generate the client certificate, the private key will remain securely stored inside the secure element which exposes an API, enabling clients to request the on chip execution of cryptographic primitive operations such as signing or ciphering.

In order to simplify deployment, secure elements are pre-personalized at manufacturing time. Each of them comes with a unique identifier (which is used also to identify the device), and these identifiers are initially provisioned in a secure element management platform along with root secrets. No software, or configuration operation needs to be performed when deploying the secure elements. They are to be considered as any other electronic component and just need to be soldered on the circuit board of the devices to secure. A bootstrap process will occur transparently upon the first connection of the device to an IP network, resulting in the possibility to remotely manage the credentials stored inside the secure element from a remote web interface.

Finally, particular attention was given to ease the use of the proposed security mechanisms by application developers that are often requesting simple to use security solutions. They want reasonable assurance that their application will provide robust data protection without having to dive into the details of the cryptographic operations. The use of TLS to protect data communications is a good example. TLS in its most common implementation involves the use of server certificates, enabling servers to be authenticated by clients. 
Clients may also be authenticated using client certificates, but the complexity involved in generating and distributing those has greatly limited the use of such authentication methods.
A very common demand from application developers is to simplify the process of obtaining the credentials they need, whatever their form. In many cases, those credentials themselves are not even handled in the developer's code, but rather passed to third party libraries or modules that the developer may be using. 

In addition to ensuring the security of communicated information in the smart grid, another cornerstone is to ensure reliable and timely delivery of the information through the used cellular networks, which is considered in the following section. 

\section{Cellular Network Performance}\label{sec:networks_performance}
The existing LTE cellular networks carry various types of traffic, e.g. mobile broadband traffic, and are expected to additionally serve the traffic originated by many IoT applications including the smart grid applications such as DSSE and demand-response. In this section the focus is on the uplink of LTE cellular network that carries the reports from the installed measurement devices (\mPMU/, PMC, and SM) towards the publish-subscriber servers.

When the (shared) existing LTE cellular networks are used to facilitate this measurement collection, from a communication performance point of view there are two possible bottlenecks that can have detrimental effects:
\begin{enumerate}[label=(\alph*)]
	\item The bottleneck in the random access phase, i.e. when a large number of smart grid devices would like to randomly or periodically transmit their measurement reports. Here, each device needs to go through the steps in the \acf{ARP}. Due to the large number of smart grid and non-smart grid devices within the LTE cell, the random access attempts to set-up the individual connections might collide resulting in failed random access attempts.
	\item The bottleneck in the communication phase (that is, after successfully finishing the random access phase) when a large number of smart grid devices would like to push their measurement data towards smart grid applications. Since many uplink messages are contending for the limited LTE uplink resources the maximum delay that some measurement reports might experience (for example due to the waiting time until a device is granted an UL transmission resource) could exceed the requirements of smart grid applications such as DSSE.
\end{enumerate}

In the following section we analyse the achievable LTE uplink delay of SM, PMC and \mPMU/ devices. Hereafter, we consider specifically the bottlenecks in the random access phase and describe a proposal for ensuring reliable random access. The analyses are based on simulation models described in references \cite{jorguseski2016lte,madueno2015massive,madueno2016assessment}.

In particular, we investigate and quantify the performance of LTE for different possible deployment scenarios in terms of the number of devices, their type, and the amount of reserved LTE uplink physical radio resources, as configured by the LTE cellular network operator for supporting the smart grid data traffic.
For the following studies, we consider the ratio R between the number of \mPMU/ devices over the number of PMC/SM devices. The PMC and SM devices are considered jointly, since their uplink traffic patterns are similar. We consider ratios of R=1/10 and R=1/3, where the former represents a scenario with moderate DER penetration where not too many \mPMU/ devices are needed, whereas the latter represents a heavy DER penetration.
The measurement report sizes from the PMC/SM and \mPMU/ devices are assumed to be 70 Bytes and 560 Bytes, respectively, as the \mPMU/ devices report more detailed power measurements.
We assume that the reporting interval of all types of measurement reports is 1~s, as motivated in sec. \ref{sec:smart_grid_monitoring}.

\subsection{Max Uplink LTE Delay for Smart Grid Data Traffic}\label{sec:max_delay}

The analysis in this section is focused on the radio part of the LTE uplink transmission, i.e., between the end-node and the LTE base station. This is because it is assumed that this is the most critical part of the end-to-end path between the measurement device and the smart-grid publish-subscribe server and applications, which are typically connected through high-speed network infrastructure as indicated in Fig.~\ref{fig:communication_architecture}. 

In LTE the uplink radio resources are organized in time-frequency blocks, also called physical resource blocks (PRBs), with duration of 0.5~ms and 12 consecutive OFDMA frequency sub-carriers. The shortest uplink radio transmission duration is 1~ms, also known as \ac{TTI}, which consists of two consecutive PRBs in the time domain. The PRB allocation per individual device and per TTI (or per block of TTIs) is done by the scheduler located in the eNB. The assumed scheduling approach for this analysis is fair fixed assignment (FFA) \cite{dimitrova2011lte} where in every TTI a device is randomly selected from a number of devices willing to transmit and it is allocated a fixed number of PRBs per device. Depending on the signal-to-Interference-plus-noise ratio (SINR) as experienced by the device on the allocated PRBs an appropriate modulation and coding scheme is selected for the transmission. This, in turn determines the amount of data (in bits) that can be transferred, and finally the number of TTIs needed to transmit the measurement report by the devices. Additionally, as there is a limited number of reserved uplink LTE radio resources (such as a limited total number of PRBs) for the transmission of the measurement reports, not all active devices in the LTE cell can begin their transmission within one TTI. As a consequence, a number of active devices have to wait until they receive an uplink transmission grant, resulting in a certain amount of waiting time. Then the maximum uplink LTE delay is the sum of the transmission time and the waiting time. For more details regarding the analysis of the maximum uplink LTE delay the reader is referred to reference \cite{jorguseski2016lte}.

In order to quantify the maximum LTE uplink delay, monte-carlo system level simulations were performed for an urban environment with increased number of total smart grid devices per LTE cell.

\begin{figure*}[bt]
    \centering
    \begin{subfigure}[b]{0.49\textwidth}
        \includegraphics[width=\textwidth]{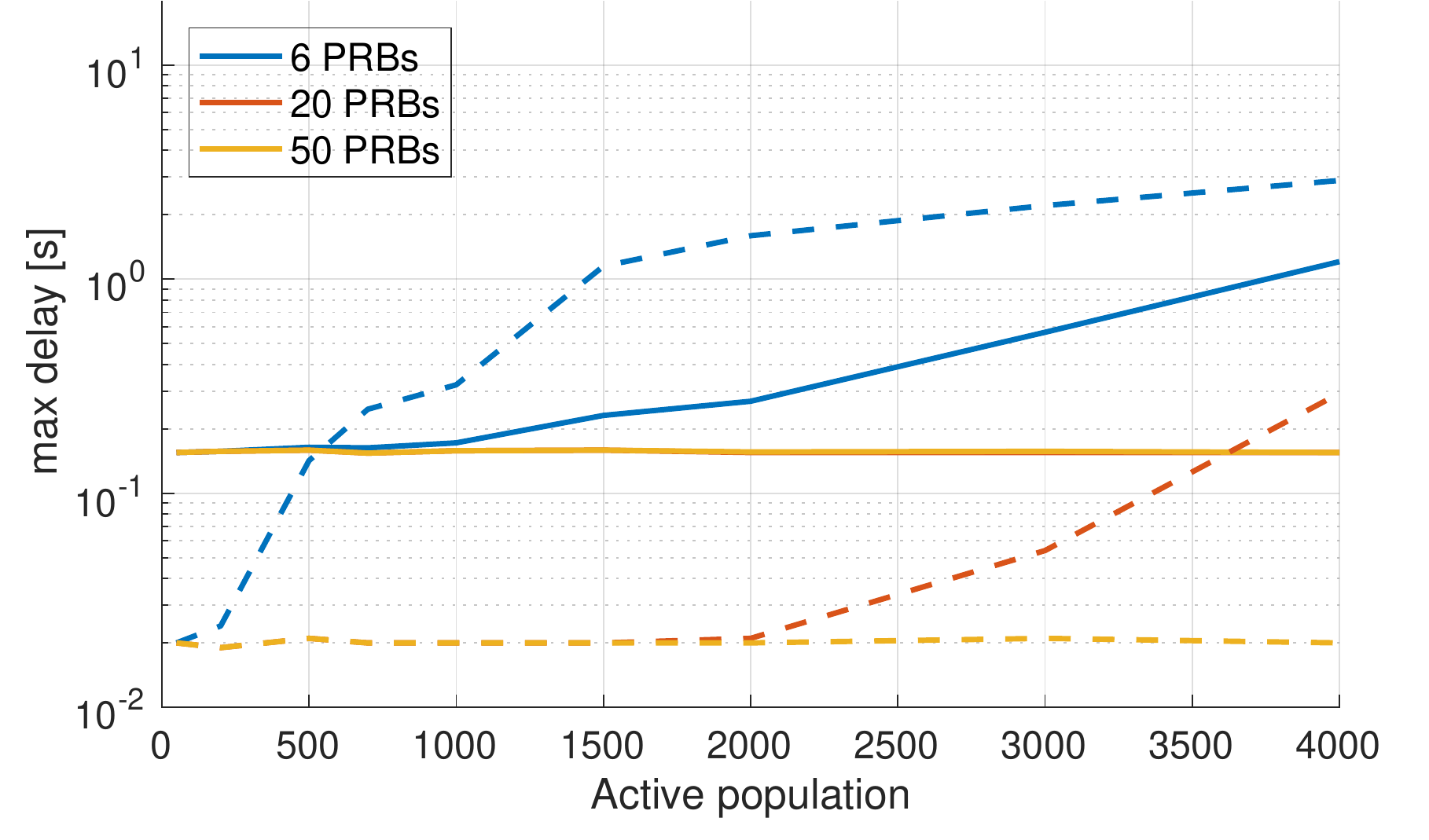}
        \caption{R=1/10 (moderate DER penetration)}
        \label{fig:wamsToSm110}
    \end{subfigure}
    ~ 
    \begin{subfigure}[b]{0.49\textwidth}
        \includegraphics[width=\textwidth]{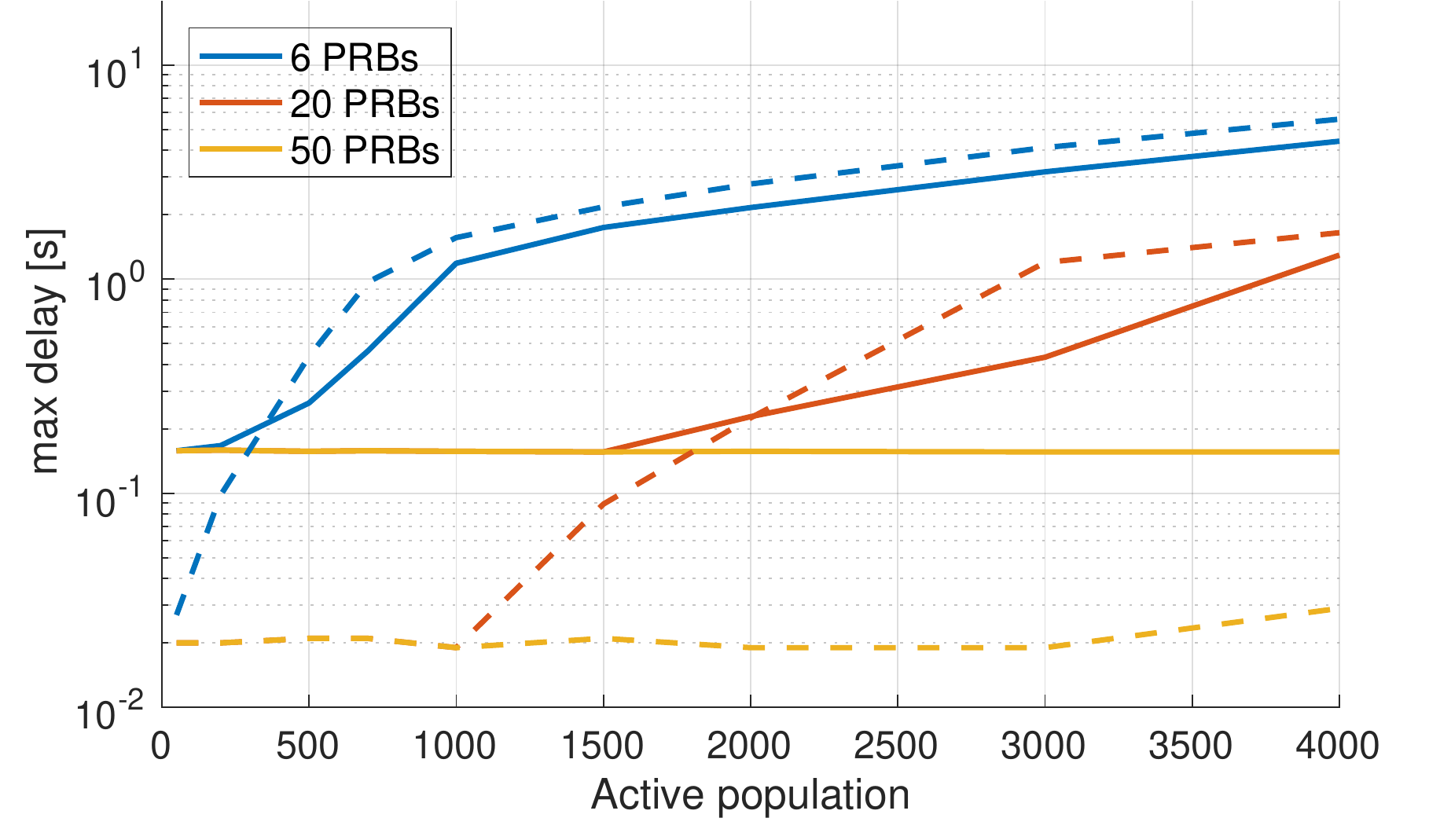}
        \caption{R=1/3 (heavy DER penetration)}
        \label{fig:wamsToSm13}
    \end{subfigure}
    \caption{PMC/SM (dashed line) and \mPMU/ (solid line) maximum delay.}\label{fig:max_delay}
\end{figure*}

In Fig.~\ref{fig:wamsToSm110} and Fig.~\ref{fig:wamsToSm13} the maximum LTE delay results are presented for R=1/10 and R=1/3, respectively, for fixed number of 2 PRBs assigned per device. It can be seen that even for very small total number of devices the maximum uplink LTE delay is about 20~ms and 200~ms, respectively. This is the intrinsic delay of transmitting the measurement report (including any re-transmissions), since for low number of devices the waiting time is practically zero. 

As the number of devices increases, the maximum uplink delay remains constant with zero waiting time, only for when the whole bandwidth (10 MHz or 50 PRBs in this case) is reserved for the smart grid data. As the number of available PRBs is decreased to 20 or 6 PRBs, which is more realistic in shared networks, the max delay rapidly increases due to the waiting time incurred at the devices until they get a scheduling grant for uplink transmission. 
The only exception here is the max delay for the \mPMU/ in Fig.~\ref{fig:wamsToSm110} where the max delays for 50 PRBs and 20 PRBs are equal and stays constant up to 4000 devices per LTE cell (i.e. these two curves overlap).
If the maximum delay requirement for the real-time application is for instance 1~s then in order to achieve this requirement for e.g. up to 4000 devices per LTE cell, the operator is required to reserve 50 PRBs (or a whole 10 MHz LTE carrier), which might not be an economically viable solution. For a more realistic amount of reserved resources (for example 6 PRBs) the achievable maximum delay is 6s or 3s for R=1/3 or R=1/10, respectively.

\subsection{Guaranteed Reliability of Random Access in LTE}\label{sec:reliable_lte}
When a measurement device wants to transmit a report, it will need to change state from idle to connected in the LTE network, through the \ac{ARP} \cite{anton2014machine}. This procedure has several steps in which failures can occur, especially in case of preamble collisions \cite{madueno2016assessment}. The collision probability increases with the number of active devices \cite{madueno2015massive}. This means that in a traditional LTE network, a large number of accessing devices can cause unacceptable delays for the mission critical traffic of DSSE and demand-response applications, where certain reliability requirements exist. Since all random access requests are treated equally in legacy LTE, there is no way to favorise certain types of traffic.

\begin{figure}[bht]
    \includegraphics[width=\linewidth]{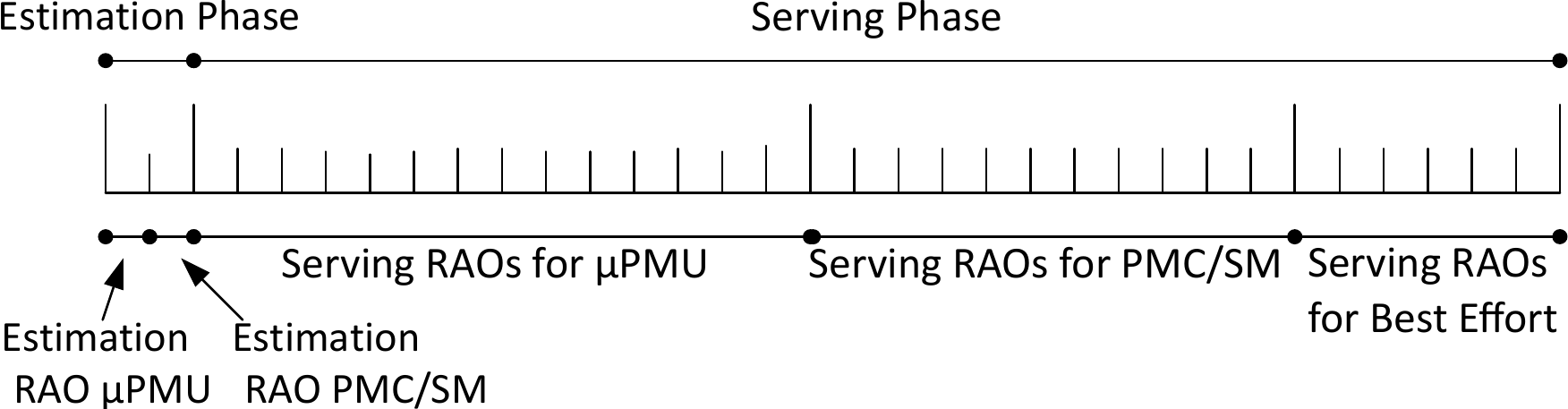}
    \caption{Proposed contention frame layout for \mPMU/, PMC/SM and best effort traffic.}
    \label{fig:contention_frame}
\end{figure}

For mission critical traffic, we propose an alternative approach to random access in LTE, which allows to reserve sufficient \acp{RAO} to ensure a certain level of success probability (reliability) in the random access procedure \cite{madueno2015massive}. Instead of serving all accessing devices equally, as in the legacy LTE ARP, the modified approach that is described in detail in reference \cite{madueno2015massive} allows to create prioritized traffic classes for which the probability of successfully accessing the network can be guaranteed. The principle of this approach is illustrated in Fig.~\ref{fig:contention_frame}, where the prioritized traffic classes \mPMU/, PMC/SM and Best Effort (ordered by most important first) that are relevant for the considered smart grid communication system are shown. Each of the \mPMU/ and PMC/SM traffic classes have a dedicated Estimation Slot in the Contention Frame, in which the corresponding devices must activate a random preamble to access the network. This enables the eNodeB to estimate the number of accessing devices and to dimension the following serving phases to satisfy the required reliability. Next, the devices will activate a preamble randomly within the corresponding serving phase, to start the ARP. Any remaining RAOs in the contention frame will be available for best effort traffic.
The duration of the contention frame is set as half of the shortest latency deadline, to ensure that the all deadlines can be fulfilled.

\begin{figure*}
    \centering
    \begin{subfigure}[b]{0.49\textwidth}
        \includegraphics[width=\textwidth]{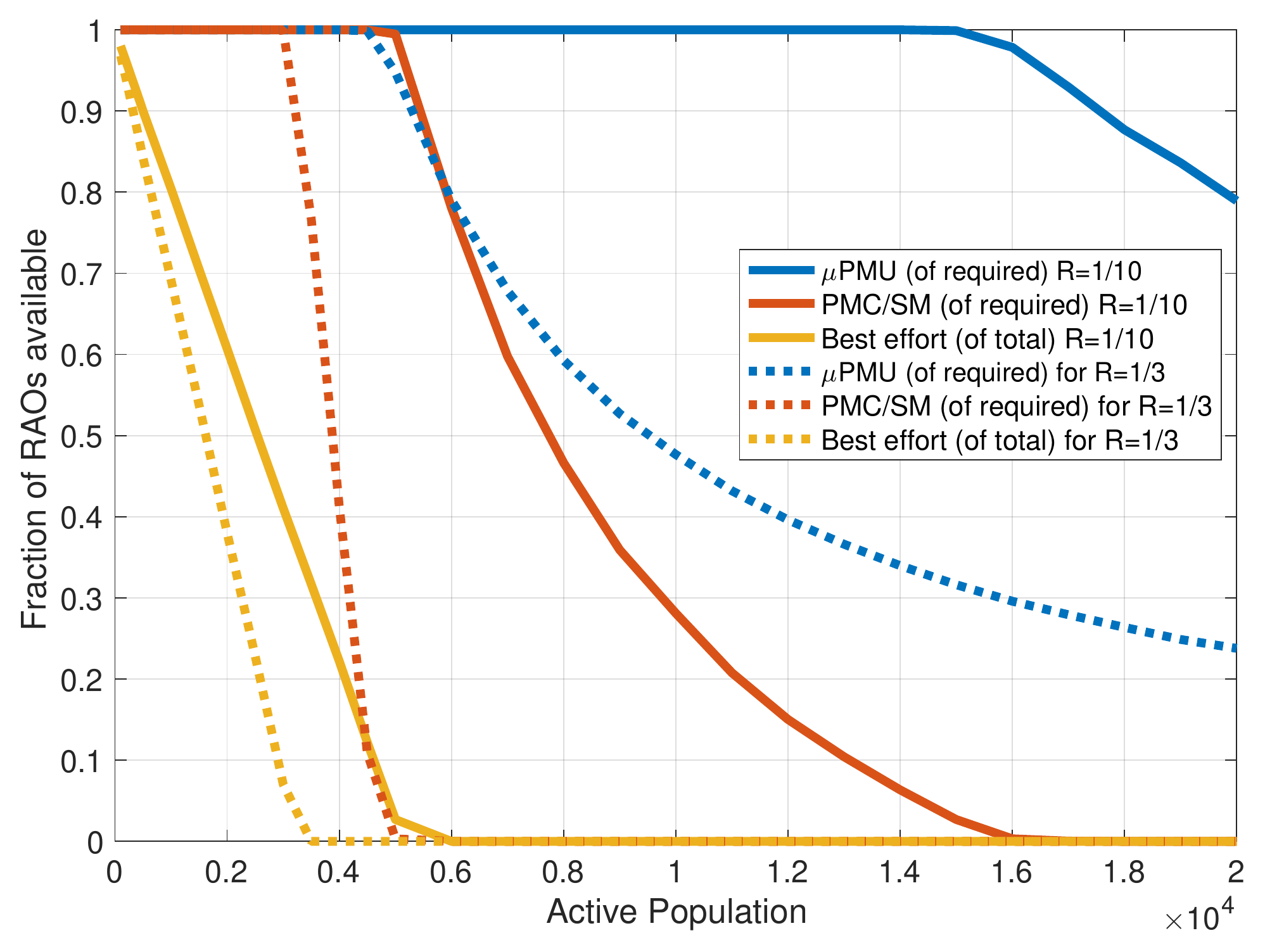}
        \caption{Availability of random access resources when the active population increases.}
        \label{fig:RACH_fraction}
    \end{subfigure}
    ~ 
    \begin{subfigure}[b]{0.49\textwidth}
        \includegraphics[width=\textwidth]{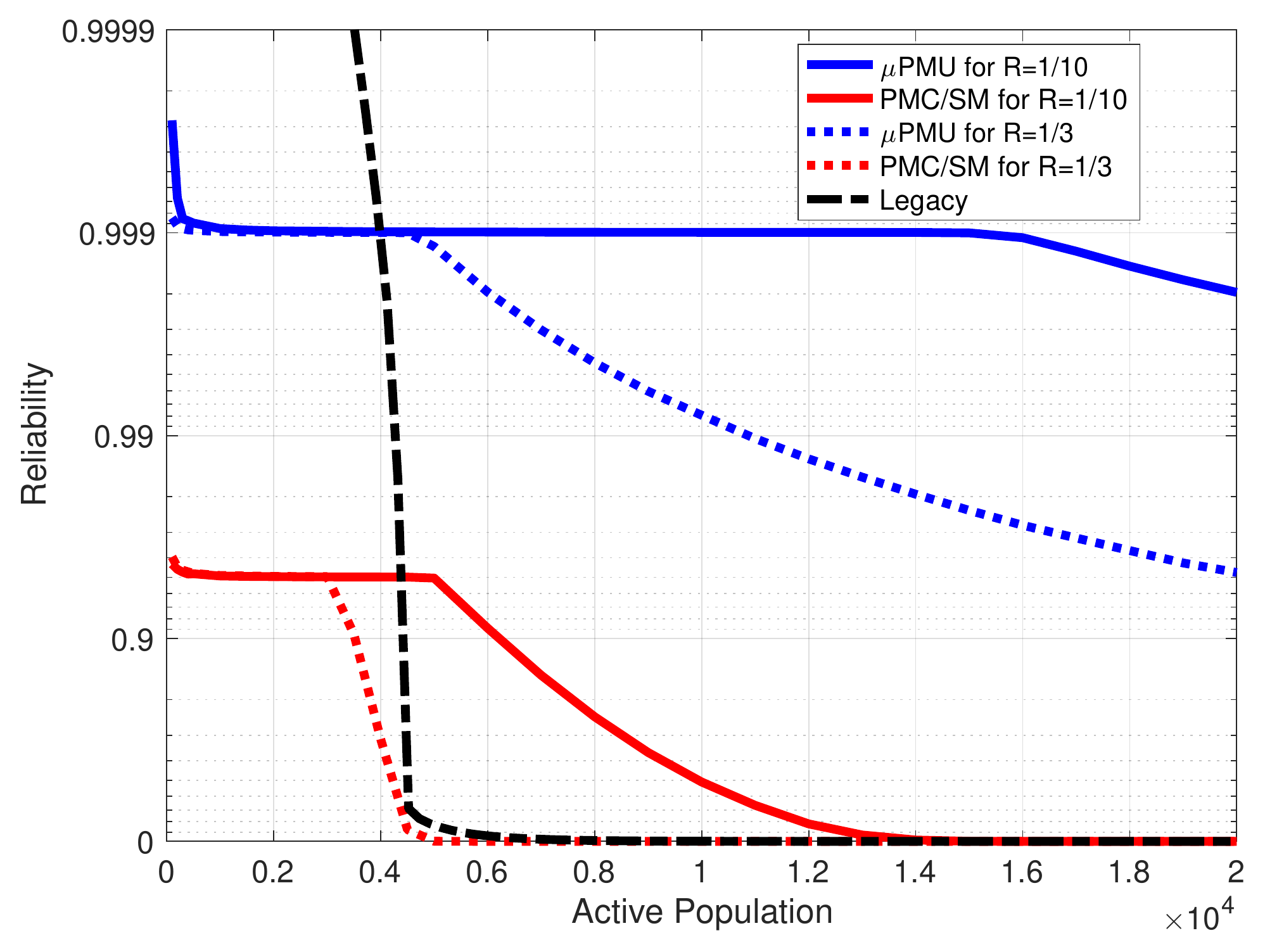}
        \caption{Achieved reliability of proposed scheme compared to legacy LTE random access.}
        \label{fig:RACH_reliability}
    \end{subfigure}
    \caption{Assumptions for result plots: 12 preambles, 1~s interval and 1~s latency deadline for both \mPMU/ and PMC/SM. Reliability: \mPMU/=99.9\%, PMC/SM=95\%.}
    \label{fig:reliable_RACH}
\end{figure*}

The allocation of resources for increasing number of \mPMU/ and PMC/SM devices in an LTE cell is shown in Fig.~\ref{fig:RACH_fraction}. The plot shows that as the number of active devices increases, first the RAOs available for best effort traffic run out. Hereafter, the RAOs for PMC/SM traffic are sacrificed to favorise the more important \mPMU/ traffic. Finally, the total number of devices becomes too large to also support \mPMU/ traffic.

For comparison, in Fig.~\ref{fig:RACH_reliability} we show the achievable reliability of the legacy LTE ARP, calculated using the collision probability model in reference \cite{madueno2016assessment}. Assuming that the arrival of  \mPMU/ and PMC/SM traffic occurs in a traditional best effort manner, the reliability of all traffic in legacy LTE will drop below the required reliability of both \mPMU/ and PMC/SM, at an earlier point than with our proposed scheme. Notice that the required reliability of \mPMU/ can be supported with the proposed scheme for 3 times as many active devices as legacy LTE for the R=1/10 scenario. This does not hold for the R=1/3 scenario where a larger fraction of devices require 99.9\% reliability, since it requires more RAOs to ensure 99.9\% reliability than 95\% reliability. 

Unfortunately, the proposed scheme for guaranteed reliability random access cannot be easily implemented in today's cellular networks, since it requires changes to the LTE protocol in both devices and eNodeB. The scheme may, however, inspire the development of \ac{MTC} protocols for the upcoming 5G standards.


\section{Conclusions and future steps} 
\label{sec:conclusion}
With the increasing penetration of \acf{DER} the smart grid needs more and deeper monitoring and control to maintain stable operation. In this paper, we have considered shared cellular LTE networks as the underlying ICT infrastructure to support the smart grid. Specifically, we highlighted the security and communication requirements such as for example end-to-end security, dynamic credential distribution, and highly reliable low latency uplink communication. Further, we outlined the solutions that were considered in the SUNSEED project for tackling the communication related challenges of ensuring successful operation of the future smart grids.

\begin{figure*}[tb]
    \includegraphics[width=\textwidth]{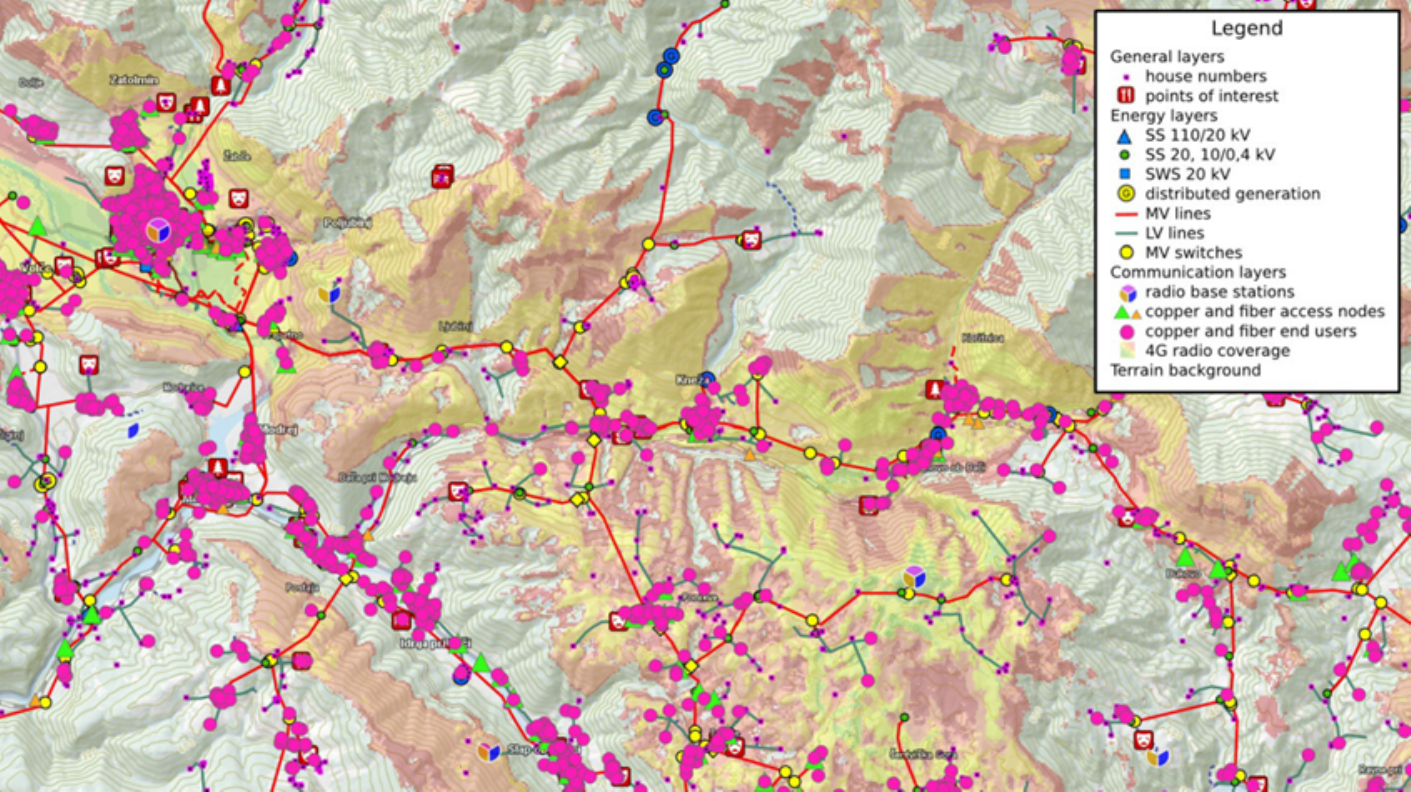}
    \caption{Kneza trial region in Slovenia including electricity and communication network infrastructures, LTE coverage, and end-users.}
    \label{fig:trial_kneza}
\end{figure*}

In the last year of the SUNSEED project (until Jan. 31st, 2017) the suitability of the LTE cellular network for facilitating the smart grid monitoring and control functions will be tested via a large field trial in Slovenia consisting of four planned areas with total number of 42 \mPMU/ devices, 5 PMC devices, 22 PLC concentrators and 563 SMs as follows:
\begin{description}
  \item [Kromberk area] with LTE coverage of 10 \mPMU/ devices, 1 PMC device, 6 PLC concentrators and 116 SMs.
  \item [Bonifika area] with LTE coverage of 7 \mPMU/ devices, 2 PMC devices, 5 PLC concentrators and 535 SMs.
  \item [Razdrto area]with UMTS coverage (due to lack of LTE coverage) of 17 \mPMU/ devices, 2 PMC devices, 7 PLC concentrators and 10 SMs.
  \item [Kneza area] as illustrated in Fig.~\ref{fig:trial_kneza} with LTE coverage of 3 \mPMU/ devices and 2 PLC concentrators as well as satellite links (due to lack of coverage of any cellular network in this mountainous region) covering 5 \mPMU/ devices, 2 PLC concentrators and 2 SMs.
\end{description}

For more details, the reader is referred to the project deliverables available via the project web-page at: \url{www.sunseed-fp7.eu}.


\section*{Acknowledgment}
This work is partially funded by EU, under Grant agreement no. 619437 ''SUNSEED''. 
The SUNSEED project is a joint undertaking of 9 partner institutions and their contributions are fully acknowledged.

\bibliographystyle{IEEEtran}

\begin{thebibliography}{10}
\providecommand{\url}[1]{#1}
\csname url@samestyle\endcsname
\providecommand{\newblock}{\relax}
\providecommand{\bibinfo}[2]{#2}
\providecommand{\BIBentrySTDinterwordspacing}{\spaceskip=0pt\relax}
\providecommand{\BIBentryALTinterwordstretchfactor}{4}
\providecommand{\BIBentryALTinterwordspacing}{\spaceskip=\fontdimen2\font plus
\BIBentryALTinterwordstretchfactor\fontdimen3\font minus
  \fontdimen4\font\relax}
\providecommand{\BIBforeignlanguage}[2]{{%
\expandafter\ifx\csname l@#1\endcsname\relax
\typeout{** WARNING: IEEEtran.bst: No hyphenation pattern has been}%
\typeout{** loaded for the language `#1'. Using the pattern for}%
\typeout{** the default language instead.}%
\else
\language=\csname l@#1\endcsname
\fi
#2}}
\providecommand{\BIBdecl}{\relax}
\BIBdecl

\bibitem{von2014every}
\BIBentryALTinterwordspacing
A.~von Meier and R.~Arghandeh, ``Chapter 34 - every moment counts:
  Synchrophasors for distribution networks with variable resources,'' in
  \emph{Renewable Energy Integration}, L.~E. Jones, Ed.\hskip 1em plus 0.5em
  minus 0.4em\relax Boston: Academic Press, 2014, pp. 429 -- 438. [Online].
  Available:
  \url{http://www.sciencedirect.com/science/article/pii/B978012407910600034X}
\BIBentrySTDinterwordspacing

\bibitem{bush2014smart}
S.~F. Bush, \emph{Smart Grid: Communication-Enabled Intelligence for the
  Electric Power Grid}.\hskip 1em plus 0.5em minus 0.4em\relax John wiley \&
  sons, 2014.

\bibitem{ozturk2013intelligent}
Y.~Ozturk, D.~Senthilkumar, S.~Kumar, and G.~Lee, ``An intelligent home energy
  management system to improve demand response,'' \emph{Smart Grid, IEEE
  Transactions on}, vol.~4, no.~2, pp. 694--701, 2013.

\bibitem{zhu2015game}
Z.~Zhu, S.~Lambotharan, W.~H. Chin, and Z.~Fan, ``A game theoretic optimization
  framework for home demand management incorporating local energy resources,''
  \emph{Industrial Informatics, IEEE Transactions on}, vol.~11, no.~2, pp.
  353--362, 2015.

\bibitem{osg2013report}
\BIBentryALTinterwordspacing
{Open Smart Grid SG-Network task force}, ``{Network Systems Requirement
  Specification v5.0 (final)},'' feb 2013. [Online]. Available:
  \url{http://osgug.ucaiug.org/}
\BIBentrySTDinterwordspacing

\bibitem{zhu2016efficient}
Z.~Zhu and Z.~Fan, ``An efficient consumption optimisation for dense
  neighbourhood area demand management,'' in \emph{IEEE International Energy
  Conference (ENERGYCON)}.\hskip 1em plus 0.5em minus 0.4em\relax IEEE, apr
  2016.

\bibitem{hardt2012oauth}
D.~Hardt, ``The oauth 2.0 authorization framework,'' 2012.

\bibitem{maler2015user}
E.~Maler, D.~Catalano, M.~Machulak, and T.~Hardjono, ``User-managed access
  (uma) profile of oauth 2.0,'' 2015.

\bibitem{etsi2015requirements}
{ETSI}, ``Ts 103†383 smart cards; embedded {UICC}; requirements specification
  (release 13),'' 2015.

\bibitem{jorguseski2016lte}
L.~Jorguseski, H.~Zhang, M.~Chrysalos, M.~Golinski, and Y.~Toh, ``{LTE} delay
  assessment for real-time management of future smart grids,'' in \emph{1st EAI
  International Conference on Smart Grid Inspired Future Technologies
  (SmartGIFT)}.\hskip 1em plus 0.5em minus 0.4em\relax EAI, 2016.

\bibitem{madueno2015massive}
G.~C. Madueno, N.~K. Pratas, C.~Stefanovic, and P.~Popovski, ``Massive {M2M}
  access with reliability guarantees in {LTE} systems,'' in
  \emph{Communications (ICC), 2015 IEEE International Conference on}.\hskip 1em
  plus 0.5em minus 0.4em\relax IEEE, 2015, pp. 2997--3002.

\bibitem{madueno2016assessment}
G.~C. Madue{\~n}o, J.~J. Nielsen, D.~M. Kim, N.~K. Pratas,
  {\v{C}}.~Stefanovi{\'c}, and P.~Popovski, ``Assessment of {LTE} wireless
  access for monitoring of energy distribution in the smart grid,'' \emph{IEEE
  Journal on Selected Areas in Communications}, vol.~34, no.~3, pp. 675--688,
  March 2016.

\bibitem{dimitrova2011lte}
\BIBentryALTinterwordspacing
D.~C. Dimitrova, J.~L. van~den Berg, G.~Heijenk, and R.~Litjens, ``{LTE} uplink
  scheduling - flow level analysis,'' in \emph{Multiple Access Communications:
  4th International Workshop, MACOM 2011}.\hskip 1em plus 0.5em minus
  0.4em\relax Springer Berlin Heidelberg, 2011, pp. 181--192. [Online].
  Available: \url{http://dx.doi.org/10.1007/978-3-642-23795-9_16}
\BIBentrySTDinterwordspacing

\bibitem{anton2014machine}
C.~Anton-Haro and M.~Dohler, \emph{Machine-to-machine {(M2M)} Communications:
  Architecture, Performance and Applications}.\hskip 1em plus 0.5em minus
  0.4em\relax Elsevier, 2014.

\end{thebibliography}
\end{document}